\documentclass[preprint,tightenlines,aps,superscriptaddress,nofootinbib,preprintnumbers,showkeys,titlepage]{revtex4-2}

\usepackage[utf8]{inputenc}
\usepackage[english]{babel}
\usepackage{graphicx}
\usepackage{physics}
\usepackage{amsmath}
\usepackage{amssymb}
\usepackage{bbm}
\usepackage{tikz}
\usepackage{xcolor}
\usepackage{slashed}
\usepackage{hyperref}
\usepackage{placeins}
\newcommand{\K}[1]{\ensuremath{\left(#1\right)}}
\newcommand{\Ke}[1]{\ensuremath{\left[#1\right]}}
\newcommand{\M}{\ensuremath{M_\Lambda}}
\allowdisplaybreaks

\begin{document}

\title{Pionic Final State Interactions and the Hypertriton Lifetime}
\author{F. Hildenbrand}
\email{f.hildenbrand@fz-juelich.de}
\affiliation{Institut für Kernphysik, Institute for Advanced Simulation and J\"ulich Center for Hadron Physics, Forschungszentrum J\"ulich, 52425, J\"ulich, Germany
}

\author{H.-W. Hammer}
\email{Hans-Werner.Hammer@physik.tu-darmstadt.de}
\affiliation{Institut für Kernphysik, Technische Universit\"at Darmstadt,
64289 Darmstadt, Germany}
\affiliation{ExtreMe Matter Institute EMMI and Helmholtz Forschungsakademie Hessen für FAIR (HFHF), GSI Helmholtzzentrum
für Schwerionenforschung GmbH, 64291 Darmstadt, Germany}

\date{\today}

\begin{abstract}
We analyze the contribution of pionic final state interactions (FSI) in the weak decay of the hypertriton. Focusing on the $^3$He channel, we find a contribution of the pionic FSI  of the order of $18\%$. Assuming a fixed value for the branching ratio $R_3$ for the decay width into $^3$He over the decay width into $^3$He and $pd$ final states, we find values for the hypertriton lifetime that are consistent with the world average as well as recent measurements by the ALICE Collaboration.
\end{abstract}
\maketitle
\section{Introduction}
The properties of hypernuclei provide an ideal playground to test the strong interaction in the strange sector.  For example, hyperons open a window onto the interior of nuclei since they are not constrained by the Pauli principle. Moreover, they are expected to play an important role in neutron stars \cite{Weber:2004kj}.  Hyperon-nucleon interactions based on chiral effective field theory (EFT) have been derived in Refs.~\cite{Polinder:2006zh,Haidenbauer:2013oca,Haidenbauer:2023qhf}. For light hypernuclei, a description in pionless EFT is also possible~\cite{Hammer:2001ng,Ando:2015fsa,Contessi:2018qnz,Hildenbrand:2019sgp}. An overview of theoretical and experimental advances in hypernuclear physics was given by Gal and collaborators~\cite{Gal:2016boi}. A comprehensive collection  of hypernuclear data can be found at~\cite{HypernuclearDataBase}. Here we focus on the few-body physics of hypernuclei. Since there are no bound hyperon-nucleon and hyperon-hyperon systems, the simplest hypernucleus is the hypertriton, a three-body bound state of a proton, a neutron, and a $\Lambda$ hyperon. It plays the same role as the "deuteron" in standard nuclear physics and provides a gateway to understand nuclear physics in the strangeness sector. 

The properties of the hypertriton have been calculated using phenomenological
interaction models as well as effective field theories (see, e.g., Refs.~\cite{Congleton:1992kk,Hammer:2001ng,Wirth:2014apa,Gal:2018bvq,Hildenbrand:2019sgp,Le:2019gjp,Perez-Obiol:2020qjy,Hildenbrand:2020kzu}).
Furthermore, first lattice QCD calculations of light hypernuclei have become available for unphysical pion masses~\cite{Beane:2012vq}. 
Since the $\Lambda$ separation energy of the hypertriton, $B_\Lambda$,
is small compared to the binding energy of the deuteron, $B_d \approx 2.2$ MeV,
the hypertriton can be viewed as a $\Lambda d$ bound state at low resolution.
The most frequently cited value for this separation energy is $B_\Lambda=(0.13\pm 0.05)$ MeV \cite{Juric:1973zq},
resulting in a large separation of the $\Lambda$ from the deuteron of about $10$ fm \cite{Hildenbrand:2019sgp}.
The recent value from ALICE~\cite{ALICE:2022sco},
$B_\Lambda=(0.102\pm 0.063\,({\rm stat.}) \pm 0.067\, ({\rm syst.}))$ MeV,
is slightly smaller but fully compatible.

Experimentally, the hypertriton lifetime presents a puzzle. Old emulsion experiments give a very broad range of values
ranging from 100 ps up to 280 ps~\cite{Block:1242347,Keyes:1968zz,Phillips:1969uy,Keyes:1970ck,Bohm:1970se,Keyes:1974ev}.
Newer heavy ion experiments, tend to lie significantly below the free $\Lambda$ lifetime of about 260~ps~\cite{Abelev:2010rv,Rappold:2013fic,Adam:2015yta,Adamczyk:2017buv}. However, more recent results favor slightly larger values. The world average for the hypertriton lifetime $237^{+10}_{-9}$~ps \cite{HypernuclearDataBase} today is mostly driven by the precise measurement of the ALICE collaboration in 2022 \cite{ALICE:2022sco}, $\tau =(253 \pm 11 \,({\rm stat.})\pm 6 \,({\rm syst.}))$~ps,  which contributes to $53\%$ and is fully compatible with the free $\Lambda$ lifetime, $\tau_\Lambda=(263\pm2)$ ps from the particle data group (PDG) \cite{ParticleDataGroup:2022pth}. A recent measurement of ALICE \cite {ALICE:2023ecf} also suggests a slightly smaller value for the free $\Lambda$ lifetime, $\tau_\Lambda=(261.07 \pm 0.37\,({\rm stat.})\pm 0.72\,({\rm syst.}))$~ps,  which is still in agreement with the PDG value.

Since the branching ratio $R_3=\Gamma_{^3{\rm He}}/(\Gamma_{^3{\rm He}}+\Gamma_{pd})$ for the decay width into $^3$He over the decay width into $^3$He and $pd$ is known to be in the range $R_3=0.3-0.4$ \cite{Block:1242347,Keyes:1970ck,Keyes:1974ev,Block:1963,Adamczyk:2017buv} and the contribution from the break up into three nucleons is small~\cite{Kamada:1997rv}, one can directly relate decay channels to the complete decay width of the hypertriton. In contrast to the direct calculation of both channels as done, e.g.,  in  Ref.~\cite{Hildenbrand:2020kzu}, however, one loses the predictive power for the branching ratio. In addition, new preliminary results of the STAR collaboration suggest a slightly smaller value, $R_3=0.272 \pm 0.030\text{(stat.)}\pm 0.042(\text{syst.)}$, which is still consistent with the identification of the hypertriton ground state as $J=1/2^+$\cite{Leung:2022sey}. 

In the recent theoretical and experimental investigations, the influence of final state interactions of the produced pion with other decay products on the hypertriton lifetime has been discussed. The purely kinematical argument that the large pion momentum  should lead to a small final state interaction seems valid. Therefore pionic final state interactions were often neglected in earlier calculations. An explicit treatment of P\'erez-Obiol et al.~\cite{Perez-Obiol:2020qjy} by distorting the outgoing pion waves leads to a contribution of the order of $10\%$. Moreover, a treatment of the pions using optical potentials for the hypertriton was done in Ref.~\cite{Cheon:1994cz}. Both calculations come to the result that there is a small but relevant contribution to the hypertriton lifetime.

In our previous work on the hypertriton lifetime~\cite{Hildenbrand:2020kzu}, which neglected the substructure of the deuteron, we relied on the kinematical suppression argument to neglect the pionic contribution, as a $10\%$ effect would come with other higher-order corrections from the hypertriton structure. Given the importance of the hypertriton lifetime puzzle, a reevaluation of this assumption in the power counting of the effective theory is in order. In the present work, we explicitly calculate the pionic final state interactions within the framework of~\cite{Hildenbrand:2020kzu} and discuss the impact on the hypertriton lifetime. 

The FSI calculation of P\'erez-Obiol~\cite{Perez-Obiol:2020qjy} et al. is purely done for the trinucleon-pion final state, and we will follow their strategy. We limit our FSI calculation to the $^3$He$\,\pi^-$ channel and use the known branching ratio $R_3$  as well as the $\Delta I=1/2$ rule to connect to the $^3$H$\,\pi^0$ as well as the  $p\,d\,\pi^-$ and $n\,d\,\pi^0$ breakup channels. Pionic final state interactions can be treated in this channel in the most straightforward way since only the trinucleon-pion interaction contributes. For the breakup into a deuteron and a nucleon a direct calculation of the pionic final state interaction is much more challenging since one has to include the multiple scattering series for the contributions of  nucleon-pion and deuteron-pion scattering. 

On the experimental side, the data for $^3$He$-\pi^-$ scattering is sparse. Experimental data as well as calculations for the scattering length are available (see, e.g., Refs.~\cite{Abela:1977mj,Mason:1980vg,Schwanner:1984sg,Baru:2002cg,Liebig:2010ki}). While there are also cross section data at finite energy \cite{Falomkin:1972fk,Albu:1982tt,Fournier:1984ex}, an actual phaseshift analysis is not available. This is problematic since the outgoing pion momentum in the decay is fixed at a rather large momentum of $k\sim 114$ MeV and hence a parametrization of the scattering amplitude by the scattering length only is not sufficient. In addition, the off-shell T-matrix would be necessary for a complete treatment. The problems and implications of this situation will be discussed below.

The paper is structured as follows: We start with a short recap of the most important features of the decay of the hypertriton into a trinucleon ($^3$He/$^3$H) state and a pion in Sec.~\ref{sec:chan}. In Sec.~\ref{sec:FSI}, we also investigate two different types of FSI approximations and their calculation, before discussing our result and its implications on the hypertriton lifetime in Sec.~\ref{sec:res}.
\section{Trinucleon channel with Final State Interactions}\label{sec:chan} 
We start be reviewing the key parts of the calculation for the trinucleon channel carried out in Ref.~\cite{Hildenbrand:2020kzu}. The contributions in the triton channel can be related to the helium channel using the $\Delta I =1/2$ rule. In the following, we thus focus on the helium channel. Because we only have two outgoing particles, energy and momentum conservation leads to a fixed pion momentum $k$. The final decay amplitude reads
\cite{Hildenbrand:2020kzu}:
 \begin{align}\label{eq: heliumw}
	\Gamma_{{^3_\Lambda\text{He}}}=\frac{G_F^2M_\pi^4}{\pi}\frac{k M_{{^3\text{He}}}}{M_{{^3\text{He}}}+\omega_{k}}\bar{Z}_{^3_\Lambda \text{H}}(B_\Lambda)\bar{Z}_{{^3\text{He}}}\K{B_{{^3\text{He}}}}\K{A_\pi^2+\frac{1}{9}\K{\frac{B_\pi}{\M+m}}^2k^2}\abs{I_q\K{k,B_\Lambda}}^2\,,
\end{align}
where $G_F$ is the Fermi coupling constant, $I_q$ is the result of a loop integral, and $A_\pi$, $B_\pi$ are the baryonic decay coupling constants. The wave function normalizations of $^3$He and the hypertriton are given by $\bar{Z}_{{^3\text{He}}}$ and $\bar{Z}_{^3_\Lambda \text{H}}$, respectively. Moreover, the energy of the outgoing pion is  $\omega_{k}=\sqrt{M_\pi^2 +k^2}$ with $k$ a fixed momentum determined by mass differences and the $\Lambda$ separation energy $B_\Lambda$. For further details see Ref.~\cite{Hildenbrand:2020kzu}.\footnote{In practice, we calculate the decay matrix element for the neutral channel and use the $\Delta I =1/2$ rule to recover the charged channel to simplify the calculation. The difference to a direct calculation is negligible \cite{Hildenbrand:2020kzu}.}

\begin{figure}[htp]
\includegraphics[width=0.9\textwidth]{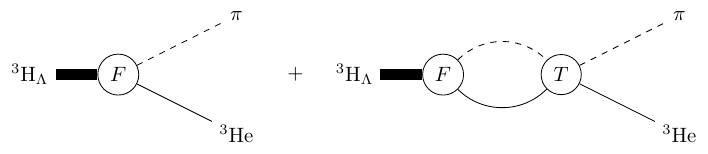}
\caption{Amplitude for the weak decay of the hypertriton into $^3$He and a pion: direct decay (left panel) and decay with pionic final state interactions (right panel). The amplitude $F$ is the weak decay amplitude, while amplitude $T$ describes the strong final state interaction\label{fig:pross}}
\end{figure}

\label{sec:FSI}
The two contributions to the hypertriton decay are depicted in Fig.~\ref{fig:pross}. Our treatment of FSI assumes that possible contributions from the three-body cut, which arises when the pion, nucleon and deuteron are on-shell, are small. A full calculation of the FSI contribution (right panel of Fig.~\ref{fig:pross}) is beyond our reach because of the lack of a phase shift analysis for $^3$He-pion scattering. We therefore establish different approximations for the loop momentum in the rescattering contribution and discuss their implications for the hypertriton lifetime.

The evaluation of the spin structure of the final state interaction is straightforward. Since the pion does not carry spin no additional spin structure is obtained, i.e., the spin structure generated by the weak vertex is inherited from the problem without pionic FSI. The two amplitudes for direct decay, $\mathcal{M}_{\slashed{\text{FSI}}}$, and decay with pionic FSI, $\mathcal{M}_{\text{FSI}}$, shown in Fig.~\ref{fig:pross} must be added coherently,
\begin{equation}
    \Gamma_{{^3\text{H}}}\sim
    \left|\mathcal{M}_{\slashed{\text{FSI}}}+\mathcal{M}_{\text{FSI}}\right|^2\,.
\end{equation}
The first diagram simply gives the direct decay amplitude
$\mathcal{M}_{\slashed{\text{FSI}}}=F(k)$. Defining \mbox{$E=k^2/(2\mu)$} as the kinetic energy of the $^3$He and pion in the final state with 
$\mu=M_\pi M_{^3{\rm He}}/(M_\pi +M_{^3{\rm He}})$,
the contribution of the second diagram in Fig.~\ref{fig:pross} can be written as:
\begin{subequations}
\begin{align}
\mathcal{M}_{\text{FSI}}&=2\mu\int\frac{\dd[3]{q}}{\K{2\pi}^3}F\K{q}\frac{1}{k^2 -q^2+i\epsilon}T\K{k}\\
&= \frac{\mu}{{\pi}^2}T\K{k}\int\dd{q}q^2 F\K{q}\Ke{\frac{\mathcal{P}}{k^2-q^2}-i\pi\delta\K{k^2-q^2}}\,.
\end{align}
\end{subequations}
Here 
\begin{equation}
    T(k) = -\frac{2\pi}{\mu} \frac{1}{k\cot\delta-ik}\,,
\end{equation}
is the elastic $\pi^- -^3$He scattering amplitude, which describes the final state interactions.\footnote{This amplitude is taken on-shell to simplify the following analysis. Any off-shell dependence could be absorbed in the definition of the typical momentum in Eq.~(\ref{def:off-shell}).}
Note that the dependence of the amplitudes $F$ and $T$ on the energy $E$ has been suppressed.
The principal value integral now determines a momentum $\bar{k}$. This momentum is the characteristic momentum of the decay amplitude $F$. Defining
\begin{align}
\int\dd{q}q^2F\K{q}\frac{\mathcal{P}}{k^2-q^2}\equiv-F\K{k}\bar{k}\frac{\pi}{2}\,,
\end{align} 
where a factor of $\pi/2$ has been pulled out of $\bar{k}$ for convenience, we obtain
\begin{align}
\mathcal{M}_{\text{FSI}}=\frac{1}{k\cot{\delta}-ik}F\K{k}\Ke{i k+\bar{k}}\,.
\label{def:off-shell}
\end{align}
The standard on-shell approximation for the 
$\pi^- -^3$He intermediate state 
can be obtained by simply discarding the contribution of the principal value integral, i.e. setting $\bar{k}=0$.
(See, e.g., Ref.~\cite{Sibirtsev:2003db} for an application to $\eta -{}^3$He scattering.)
In this case, we obtain for the total 
contribution of the two diagrams in 
Fig.~\ref{fig:pross}:
\begin{equation}
 \left|\mathcal{M}_{\slashed{\text{FSI}}}+\mathcal{M}_{\text{FSI}}\right|^2 =\left| F(k)\K{1+\frac{ik}{k\cot\delta-ik}} \right|^2 = F^2\K{k} \frac{\K{k\cot\delta}^2}{\K{k\cot\delta}^2+k^2}\equiv F^2\K{k}P_{\slashed{E}}\K{\delta}\, ,
 \label{eq:on-shellT}
\end{equation}
where we have assumed that the decay amplitude $F(k)$ is real. From the form of $P_{\slashed{E}}$ it is obvious that this approximation can only lead to a reduction of the total amplitude.

Including the contribution of the principal value integral, we obtain
\begin{equation}
 \left|\mathcal{M}_{\slashed{\text{FSI}}}+\mathcal{M}_{\text{FSI}}\right|^2 =\left| F(k)\K{1+\frac{ik+\bar{k}}{k\cot\delta-ik}} \right|^2 = F^2(k)\frac{\K{k\cot{\delta}+\bar{k}}^2}{\K{k\cot{\delta}}^2+k^2}\equiv F^2\K{k}P_E\K{\delta}\,,
 \label{eq:principalT}
\end{equation}
which depends on the characteristic momentum $\bar{k}$.

\begin{figure}[htp]
\includegraphics[width=0.45\textwidth]{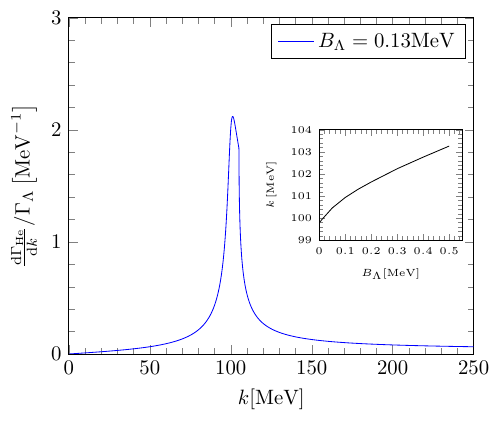}
\includegraphics[width=0.45\textwidth]{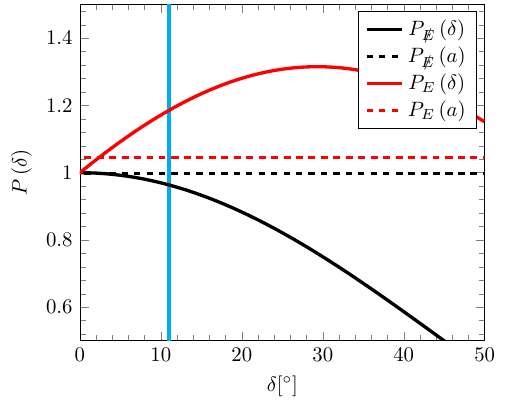}
\caption{
Left panel: Square of the decay amplitude before enforcing energy conservation for a $\Lambda$ separation energy $B_\Lambda=0.13$ MeV. The inset shows the weak dependence of the peak position $k$ on the value of $B_\Lambda$. Right panel: results for different approximations for the FSI factor $P\K{\delta}$. The black lines represent FSI factors calculated by the on-shell method  
while the red ones take loop effects into consideration. The dashed lines represent a calculation where the $\pi-$He T-Matrix is replaced by its threshold value. The cyan line is the value of the phase shift obtained by an analysis of the data in Ref.~\cite{Fournier:1984ex}. \label{fig:F}}
\end{figure}

In order to estimate $\bar{k}$ we evaluate the
integrand leading to Eq.~\eqref{eq: heliumw} without delta distribution enforcing energy conservation, cf. Eq.~(17) in Ref.~\cite{Hildenbrand:2020kzu}.  The result of this calculation is depicted in Fig.~\ref{fig:F}. It shows a strong peak around $k\sim100$ MeV, resulting in a value $\bar{k}=2k/\pi\sim 64$ MeV, which is a good estimate for the characteristic momentum $\bar{k}$. The dependence on the $\Lambda$ separation energy $B_\Lambda$ is shown in the inset. As expected the dependence is very weak, since the width goes to zero as $\sqrt{B_\Lambda}$ \cite{Hildenbrand:2020kzu}. For $B_\Lambda\leq 0.5$ MeV, it is of  the order of $3\%$ or less and hence can be neglected. In principle, similar characteristic momenta can also be defined for the three- and four-body decays, as done by Kamada et al.~\cite{Kamada:1997rv}.
In the next section, we use Eqs.~(\ref{eq:on-shellT}, \ref{eq:principalT}) together with the available experimental information on $\pi^- -^3$He scattering to estimate the contribution of pionic FSI to the hypertriton lifetime.

\section{Results and discussion}\label{sec:res}
The scattering data  for pion scattering on the trinucleon is very sparse.  Cross section data for charged pion scattering on $^3$He is available at higher energies (see, for example, Ref.~\cite{Albu:1982tt}), while data for lower energies can be found in Ref.~\cite{Fournier:1984ex}. However, we are not aware of a phase shift analysis for $^3$He$-\pi^\pm$ scattering. The scattering length for charged pion scattering off $^3$He, in contrast, is well explored in X-ray experiments \cite{Abela:1977mj,Mason:1980vg,Schwanner:1984sg} as well as in theoretical calculations~\cite{Liebig:2010ki,Baru:2002cg}. The resulting value is small, $a\approx 50\cdot 10^{-3}M_\pi^{-1}$.
However, replacing the phase shift with its threshold value is not expected to be a good approximation for the hypertriton decay as the FSI happens far away from threshold.

Before considering the available $^3$He$-\pi^\pm$ scattering data, we estimate the effect of pionic FSI based on general considerations. 
Results for the correction factors $P_{\slashed{E}}$ and $P_{{E}}$ are shown in Fig.~\ref{fig:F} as a function of the phase shift $\delta$. The on-shell FSI correction factor satisfies $P_{\slashed{E}}\K{\delta}\sim\abs{\cos(\delta)}^2$ and thus can only lead to a reduction of the decay width.
If the phase shift is approximated by the threshold value determined by the scattering length, denoted by $P_{\slashed{E}}(a)$ and $P_{{E}}(a)$ in Fig.~\ref{fig:F}, the correction is very small. Therefore, we focus now on the more realistic FSI factor $P_{E}\K{\delta}$, which includes both loop effects and information about the phase shift at finite energy.
Even small phases, such as $5^\circ$ allow correction of the order $ 10\% $ in the hypertriton width. This would make it of the same order as, e.g.,  next-to-leading order hypernuclear contributions which can be estimated from the $\Lambda d$ effective range and scattering length as $r_{\Lambda d}/a_{\Lambda d}\approx 0.1$  at $B_\Lambda = 0.13$ MeV \cite{Hildenbrand:2019sgp}. Since our previous result for the hypertriton lifetime~\cite{Hildenbrand:2020kzu} is consistent with the free $\Lambda$ lifetime, such a correction would lead to lifetimes larger than $75\%$ of the free one, $\tau_\Lambda$. A $10\%$ correction as obtained by Perez-Obiol \cite{Perez-Obiol:2020qjy}, 
would lead to a lifetime of around 0.9$\tau_\Lambda$. The maximum enhancement of about $30\%$ is reached for $\delta\approx 30^\circ$.

In the next step, we explicitly use the differential cross section 
data by Fournier et al. at $T^{\text{lab}}_\pi=(45.1\pm1.0)$ MeV \cite{Fournier:1984ex}. This pion kinetic energy  corresponds to a center-of-mass momentum $k\approx117$~MeV which is close to the fixed momentum $k\approx 114$~MeV of the outgoing pion in the hypertriton decay and the peak momentum $k\approx 100$ MeV from the left panel of Fig.~\ref{fig:F}. A fit of the theoretical expression for the differential cross section to the experimental data leads to a scattering phase shift $\delta_0\approx 11^\circ$. (Further details of this analysis are given in Appendix \ref{app_He}.) This value is indicated by the cyan line in Fig.~\ref{fig:F}. Using this estimate, we obtain an enhancement of $18\%$ using the correction factor $P_E\K{\delta}$. 

\begin{figure}[htp]
\includegraphics[width=0.9\textwidth]{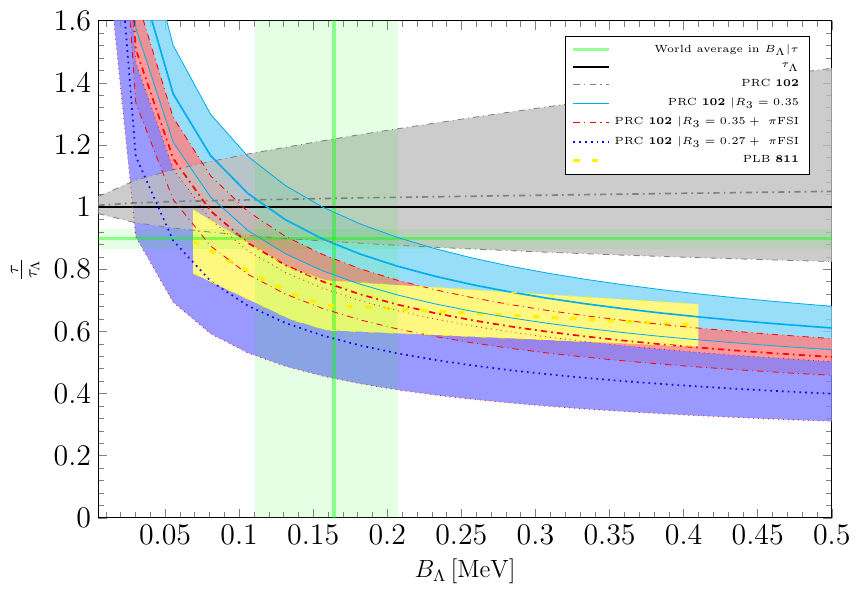}
\caption{Lifetime of the hypertriton $\tau$ relative to the free $\Lambda$ lifetime $\tau_\Lambda$. 
The grey band represents the result obtained in Ref.~\cite{Hildenbrand:2020kzu}. In addition, we show the results using a fixed branching ratio $R_3$ to calculate the total width with and without pionic final state interactions, similar to the approach  in Ref.~\cite{Perez-Obiol:2020qjy}, which is given in yellow for reference. The green areas give the world average for the hypertriton lifetime as well as the $\Lambda$ separation energy according to Ref.~\cite{HypernuclearDataBase}. In order to keep the plot readable we show EFT uncertainty bands only for the results of Ref.~\cite{Hildenbrand:2020kzu}.  \label{fig:lifetime}}
\end{figure}

We now focus on the consequences for the hypertriton lifetime $\tau$ as a function of the $\Lambda$ separation energy $B_\Lambda$. We again consider two scenarios to obtain the full decay width from the width for decay into a trinucleon and a pion: (i) we use the branching ratio $R_3$ as calculated in Ref.~\cite{Hildenbrand:2020kzu} without pionic FSI and (ii) we use the experimental value for $R_3$ as input. 
Using scenario (i) we obtain a lifetime $\tau\approx 0.84\tau_\Lambda$ which is almost independent of the
$\Lambda$ separation energy $B_\Lambda$. Note, however, that this result is not valid in the limit $B_\Lambda \to 0$ where $\tau$ should approach the free $\Lambda$ lifetime $\tau_\Lambda$.
In scenario (ii), we use the experimental branching ratio $R_3$ to acquire the full lifetime of the hypertriton. The result of this procedure is shown in Fig.~\ref{fig:lifetime}. We apply the  average of the branching ratios $R_3$ obtained in Refs.~\cite{Block:1242347,Keyes:1970ck,Keyes:1974ev,Block:1963,Adamczyk:2017buv} as well as the new proposed value by STAR~\cite{Leung:2022sey} with and without the inclusion of pionic FSI. For comparison, we show the calculation by P\'erez-Obiol \cite{Perez-Obiol:2020qjy} (yellow region), the world averages for $B_\Lambda$ and $\tau$, and the result of our previous calculation \cite{Hildenbrand:2020kzu}. In order to not clutter the plot any further, we only draw the EFT uncertainty band for the original result. 

Note that the lifetime goes to infinity as $B_\Lambda\rightarrow 0$ since $\Gamma_{{^3\text{He}}}\rightarrow 0$. This is clearly an artefact of assuming a constant value of $R_3$. It is not present in the calculation of Ref.~\cite{Hildenbrand:2020kzu} where $R_3$ also vanishes in this limit. Using the experimental value of $R_3$, the results of Refs.~\cite{Perez-Obiol:2020qjy,Hildenbrand:2020kzu} are in agreement with each other within uncertainties. Moreover, the inclusion of pionic final state interactions in both scenarios (i) and (ii) is consistent with the world averages of $B_\Lambda$ as well as the lifetime. Smaller binding energies or lifetimes closer to the free $\Lambda$ lifetime are also compatible. Higher binding energies seem to have more tension with the current average lifetime.

In this paper, we discussed different methods to include pionic final state interactions in our calculation of the hypertriton lifetime in the EFT approach of Ref.~\cite{Hildenbrand:2020kzu}. Following the strategy of Ref.~\cite{Perez-Obiol:2020qjy}, we calculate the decay into $^3$He and a $\pi^-$ explicitly and recover the full decay width by using the known branching ratio $R_3$ for the decay width into $^3$He over the decay width into $^3$He and $pd$ final states. The contribution of the neutral channels is obtained from the phenomenological $\Delta I = 1/2$ rule.
Using the on-shell approximation for the FSI correction factor leads to small corrections and an increased lifetime. Lifting this restriction and including loop effects  yields contributions similar in size as found in Ref.~\cite{Perez-Obiol:2020qjy} in a pion distorted wave approach. We obtain an ehancement  of the width by $18\%$, which is consistent with the original power counting of the EFT in Ref.~\cite{Hildenbrand:2020kzu}. Thus the contribution of pionic FSI must be included at next-to-leading order. Applying a fixed branching ratio allows predictions for the total width as a function of the $\Lambda$ separation energy. The results obtained in this way are consistent with the current experimental averages of the hypertriton lifetime and binding energy.
\FloatBarrier
\begin{acknowledgments}
We thank Christoph Hanhart, Deborah R\"onchen, and Matthias G\"obel for useful discussions.
This work was supported in part by the European
Research Council (ERC) under the European Union's Horizon 2020 research
and innovation programme (grant agreement No. 101018170), by the Deutsche Forschungsgemeinschaft (DFG, German Research Foundation) - Projektnummer 279384907 - SFB 1245, and by the German Federal Ministry of Education and Research (BMBF) (Grant No. 05P21RDFNB).
\end{acknowledgments}
\appendix
\section{\texorpdfstring{$\boldsymbol{\pi^-$--${}^3}$}~He scattering\label{app_He}}
\begin{figure}[htp]
\includegraphics[width=0.5\textwidth]{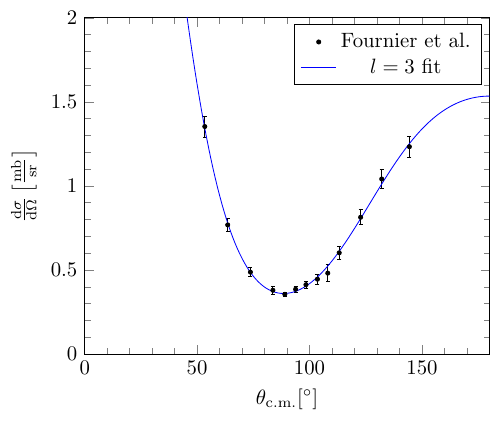}
\caption{Least squares fit of the differential cross section with $l\leq 3$ for $T^{\text{lab}}_\pi=(45.1\pm1.0)$ MeV from Fournier et al.~\cite{Fournier:1984ex}. The resulting phase shifts for elastic scattering are $\delta_0=-11.0523^\circ,\,\delta_1=-3.803^\circ,\,\delta_2=-9.115^\circ,\,\delta_3=-3.033^\circ$. \label{fig:phasefit}}
\end{figure}

As mentioned in the main text a full phase analysis for $\pi^-$--${}^3$He scattering is not currently available. However, some selected data points for the differential cross section exist at $T^{\text{lab}}_\pi=(45.1\pm1.0)$ MeV which corresponds to good agreement with the fixed outgoing center of mass momentum of $k\approx 114$ MeV
\begin{align}
    k_{\text{com}}=M_{^3\text{He}}\sqrt{\frac{T^{\text{lab}}_\pi(T^{\text{lab}}_\pi+2M_\pi)}{\K{M_{^3\text{He}}+M_\pi}^2+2T^{\text{lab}}_\pi M_{^3\text{He}}}}=(116.9\pm1.6)\,\text{MeV}\,.
\end{align}
Although the data does not allow for a full partial wave analysis, it allows for a reasonable estimation of the phase shift at the relevant energy. As a starting point we use the standard definitions of the scattering amplitude $f_k$ and cross section $\sigma$ expanded in terms of Legendre polynomials $P_l$:
\begin{align}
    f_k\K{\theta}=\sum_{l=0}^\infty \frac{1}{2ik}\K{\eta_i\exp\K{2i\delta_i}-1}P_l\K{\cos\K{\theta}}\qq{and} \sigma=\abs{f_k\K{\theta}}^2\, ,
\end{align}
where $\delta_i$  are the real parts of the phase shifts and $\eta_i$ parameterize the imaginary parts.
A first straightforward estimate for the phase shift can be obtained by assuming that the scattering process is elastic ($\eta_i=1$). 
Least square fits for the  dominant contributions stabilise at $l=3$. Higher angular momenta do not improve the reduced $\chi^2$ of the fit. The inclusion of inelastic parts in the fit does yield further insights. This is due to the high number of parameters, even for a low order fit, compared to the $13$ data points available. The result of the fit with $l=3$ is shown in Fig.~\ref{fig:phasefit}. The corresponding phase shifts for elastic scattering with $l=0,1,2,3$ at $T^{\text{lab}}_\pi=(45.1\pm1.0)$ MeV are $\delta_0=-11.0523^\circ$, $\delta_1=-3.803^\circ$, $\delta_2=-9.115^\circ$, and $\delta_3=-3.033^\circ$. 
Thus the S-wave phase shift which is relevant for the FSI factor is about $\sim 11$ degrees.
Adding inelastic parts to the fit $(\eta_i\neq1)$ does not decrease the reduced $\chi^2$. The S-Wave phase shift is stable against adding higher partial waves and inelastic parts, however, the data will be clearly overfitted in this cases. Explicit Coulomb effects are expected to be small at high energies and thus neglected. 
\FloatBarrier
\bibliography{lifetimelib.bib}
\end{document}